\documentclass[onecolumn,preprint,superscriptaddress]{revtex4-1}
\usepackage{graphicx}
\usepackage{dcolumn}
\usepackage{bm}
\usepackage{amssymb}
\usepackage[version=3]{mhchem} 
\usepackage{dsfont}
\usepackage{hyperref}
\usepackage{subfigure}
\usepackage[T1]{fontenc}
\usepackage{graphicx}
\usepackage{siunitx}

\makeatletter
\renewcommand\tagform@[1]{\maketag@@@{\ignorespaces#1\unskip\@@italiccorr}}
\makeatother

\DeclareMathAlphabet      {\mathbfit}{OML}{cmm}{b}{it}

\begin{document}
\title{Temperature Effect on Charge-state Transition Levels of Defects in Semiconductors}

\author{Shuang Qiao}
\affiliation{Beijing Computational Science Research Center, Beijing 100193, China}

\author{Yu-Ning Wu}
\affiliation{Key Laboratory of Polar Materials and Devices (MOE) and Department of Electronics, East China Normal University, Shanghai 200241, China}

\author{Xiaolan Yan}
\affiliation{Beijing Computational Science Research Center, Beijing 100193, China}

\author{\\Bartomeu Monserrat}
\affiliation{Cavendish Laboratory, University of Cambridge, Cambridge CB3 0HE, U.K.}
\affiliation{Department of Materials Science and Metallurgy, University of Cambridge, Cambridge CB3 0FS, U.K}

\author{Su-Huai Wei}
\affiliation{Beijing Computational Science Research Center, Beijing 100193, China}
\affiliation{Department of Physics, Beijing Normal University, Beijing 100875, China}

\author{Bing Huang}
\email{Bing.Huang@csrc.ac.cn}
\affiliation{Beijing Computational Science Research Center, Beijing 100193, China}
\affiliation{Department of Physics, Beijing Normal University, Beijing 100875, China}

\date{\today}

\newpage
\begin{abstract}
\textbf{Defects are crucial in determining the overall physical properties of semiconductors. Generally, the charge-state transition level $\varepsilon$$_{\alpha}$(\emph{q}/\emph{q'}), one of the key physical quantities that determines the dopability of defects in semiconductors, is temperature dependent. However, little is known about the temperature dependence of $\varepsilon$$_{\alpha}$(\emph{q}/\emph{q'}), and, as a result, almost all existing defect theories in semiconductors are built on a temperature-independent approximation. In this article, by deriving the basic formulas for temperature-dependent $\varepsilon$$_{\alpha}$(\emph{q}/\emph{q'}), we have established two fundamental rules for the temperature dependence of $\varepsilon$$_{\alpha}$(\emph{q}/\emph{q'}) in semiconductors. Based on these rules, surprisingly, it is found that the temperature dependences of $\varepsilon$$_{\alpha}$(\emph{q}/\emph{q'}) for different defects are rather diverse: it can become shallower, deeper, or stay unchanged. This defect-specific behavior is mainly determined by the synergistic or opposing effects between free energy corrections (determined by the local volume change around the defect during a charge-state transition) and band edge changes (which differ for different semiconductors). These basic formulas and rules, confirmed by a large number of state-of-the-art temperature-dependent defect calculations in GaN, may potentially be widely adopted as guidelines for understanding or optimizing doping behaviors in semiconductors at finite temperatures.}
\end{abstract}

\maketitle

\section*{I. Introduction}
Intrinsic defects and external impurities (generally denoted as defects hereafter) play a critical role in determining the physical properties of solids, \emph{e.g.}, from solar cells \cite{1,2,3} to solid-state lighting \cite{4,5} to topological phase control \cite{6,7,8} and to quantum computing \cite{9,10,11}. The defect formation energies \emph{H}$_{\emph{f}}$(\emph{$\alpha$},\emph{q}) for defect $\alpha$ at charge state \emph{q}, that determine the defect concentrations; and the charge-state transition levels $\varepsilon$$_{\alpha}$(\emph{q}/\emph{q'}), that correspond to the thermal ionization energies, are two of the most important physical quantities for all the defects in semiconductors \cite{12,13,50}. Generally, both \emph{H}$_{\emph{f}}$(\emph{$\alpha$},\emph{q}) and $\varepsilon$$_{\alpha}$(\emph{q}/\emph{q'}) are temperature dependent. Differing from the straightforward temperature dependence of \emph{H}$_{\emph{f}}$(\emph{$\alpha$},\emph{q}) \cite{12,13,14,17,18}, little is known about how temperature changes affect $\varepsilon$$_{\alpha}$(\emph{q}/\emph{q'}) in semiconductors, due to the lack of basic formulas and fundamental rules. As a result, almost all defect theories in semiconductors are built on static first-principles calculations excluding temperature effects \cite{12,13,50}.

The challenge to unravel the temperature dependence of $\varepsilon$$_{\alpha}$(\emph{q}/\emph{q'}) in theory is two-fold. Fundamentally, the standard formulas for $\varepsilon$$_{\alpha}$(\emph{q}/\emph{q'}) calculations are incomplete and do not capture the $\varepsilon$$_{\alpha}$(\emph{q}/\emph{q'}) of defects under finite temperatures. Practically, the computations of temperature-induced vibrational properties of defects in semiconductors are extremely expensive. Because of its unparalleled complexity, the temperature dependence of $\varepsilon$$_{\alpha}$(\emph{q}/\emph{q'}) in semiconductors has remained unanswered for decades i.e., we do not have any rules to predict or understand the dopability of semiconductors at finite or changing temperatures.

Differing from narrow bandgap (NBG) semiconductors (\emph{e.g.}, Si and GaAs) that usually operate under ambient environments at room temperature, wide bandgap (WBG) semiconductors (\emph{e.g.}, GaN and SiC) can operate under harsh environments with high working temperatures \cite{20,21,22,23,24}. Therefore, WBG semiconductors are an ideal platform for unique applications in aerospace, nuclear power and earth's mantle investigation, that require changing operation temperatures from extremely-low to extremely-high (0$\sim$1000 K) \cite{20,21,22,23,24,25,26,27,28}. This highlights the need to understand the evolution with temperature of defect properties in WBG semiconductors, especially of $\varepsilon$$_{\alpha}$(\emph{q}/\emph{q'}), which may be critical to improve the reliability of WBG semiconductor devices in various environments.

In this article, by deriving the basic formulas of temperature-dependent $\varepsilon$$_{\alpha}$(\emph{q}/\emph{q'}), we have established two fundamental rules for the temperature dependence of absolute and relative $\varepsilon$$_{\alpha}$(\emph{q}/\emph{q'}) in semiconductors, respectively. Based on these rules, it is found that regardless of the initial $\varepsilon$$_{\alpha}$(\emph{q}/\emph{q'}) levels at 0 K, surprisingly, the temperature-dependent behaviors of $\varepsilon$$_{\alpha}$(\emph{q}/\emph{q'}) for different defects in different types of semiconductors are rather diverse, \emph{i.e.}, it can become shallower, deeper, or even stay unchanged, mainly determined by the synergistic or opposing effects between free energy corrections and band edge changes. Importantly, we discover that the electronic and vibrational contributions to free energy corrections are both fundamentally determined by a key physical quantity $\delta$\emph{V}$_{q\rightarrow q'}$, the local volume change around the defect during the charge-state transition. Interestingly, the $\delta$\emph{V}$_{q\rightarrow q'}$ values are mainly determined by the competing effect between the local electron occupation (LEO) changes and the strength of the local lattice relaxation (LLR) around the defects. Using the state-of-art first-principles-based temperature-dependent approaches with the capacity of both high accuracy and high efficiency \cite{51}, these proposed basic formulas and fundamental rules have been thoroughly verified based on a large number of defect calculations in GaN.
\section*{II. Results and Discussion}
\subsection*{A. Basic Formulas}
Without the inclusion of temperature effects, the $\varepsilon$$_{\alpha}$(\emph{q}/\emph{q'}) of a defect $\alpha$ between the charge-states \emph{q} and \emph{q'} is given as
\begin{equation}
\varepsilon_{\alpha}(\emph{q}/\emph{q'})_{w.o.T}=\frac{E(\alpha,q')-E(\alpha,q)}{\emph{q}-\emph{q'}}-\varepsilon_{VBM}(\emph{host}),   \tag{(1)}
\end{equation}
\noindent where \emph{E}($\alpha$,\emph{q}) [\emph{E}($\alpha$,\emph{q'})] is the total energy of a supercell with defect $\alpha$ in charge-state \emph{q} [\emph{q'}] and $\varepsilon$$_{VBM}$(\emph{host}) is the valence band maximum (VBM) of the host \cite{12,13,50}. With the inclusion of temperature effects, \emph{E}($\alpha$,\emph{q}) [\emph{E}($\alpha$,\emph{q'})] in Eq.(1) is replaced by the corresponding free energy \emph{F}($\alpha$,\emph{q}) [\emph{F}($\alpha$,\emph{q'})]. After some manipulations [see Appendix A], it can be written as
\begin{equation}
\varepsilon_{\alpha}(\emph{q}/\emph{q'})[\emph{V},\emph{T}\,] = \varepsilon_{\alpha}(\emph{q}/\emph{q'})_{w.o.T} + \frac{\Delta\emph{F}^{\,el}[\emph{V},\emph{T}\,] + \Delta\emph{F}^{\,ph}[\emph{V},\emph{T}\,]}{\emph{q}-\emph{q'}} - \Delta\varepsilon_{VBM}(\emph{host})[\emph{V},\emph{T}\,].  \tag{(2)}
\end{equation}
\noindent On the right-hand side of Eq. (2), the second term represents the corrections from the free energy differences between the \emph{q} and \emph{q'} configurations induced by the electronic ($\Delta$\emph{F}$^{el}$) and vibrational ($\Delta$\emph{F}$^{ph}$) contributions, while the third term represents the correction on the VBM energy position driven by thermal expansion and electron-phonon coupling ($\Delta$$\varepsilon$$_{VBM}$ = $\Delta$$\varepsilon$$^{th}$$_{VBM}$+$\Delta$$\varepsilon$$^{ph}$$_{VBM}$). In practice, the temperature dependence of $\varepsilon$$_{\alpha}$(\emph{q}/\emph{q'}) can be understood without and with the inclusion of $\Delta$$\varepsilon$$_{VBM}$ \cite{45}, corresponding to the absolute and relative evolutions of $\varepsilon$$_{\alpha}$(\emph{q}/\emph{q'}), respectively. While the temperature dependence of the absolute $\varepsilon$$_{\alpha}$(\emph{q}/\emph{q'}) [$\varepsilon$$^{a}$$_{\alpha}$(\emph{q}/\emph{q'})] is solely determined by the free energy corrections, that of the relative $\varepsilon$$_{\alpha}$(\emph{q}/\emph{q'}) [$\varepsilon$$^{r}$$_{\alpha}$(\emph{q}/\emph{q'})] is determined by both the free energy corrections and the band edge changes. Although $\varepsilon$$_{\alpha}$(\emph{q}/\emph{q'}) is independent of the direction of charge-state transitions, to simplify our discussion, in the following we focus on the ionization process, \emph{i.e.}, |\emph{q'}|>|\emph{q}|. This assumption does not change the rules we developed.

Under the quasi-harmonic approximation (QHA), \emph{F}$^{el}$ can be written as \emph{F}$^{el}$=\emph{E}$^{th}$+\emph{E}$^{el}$-\emph{T}\emph{S}$^{el}$ \cite{13,29,30}, where the first, second, and third terms are the energy corrections induced by thermal expansion, electron-occupation change, and electronic entropy, respectively. Generally, the contributions from \emph{E}$^{el}$ and \emph{S}$^{el}$ to \emph{F}$^{el}$ are negligible under reasonable temperatures in semiconductors \cite{12,31}. Therefore, we focus on the \emph{E}$^{th}$ term in \emph{F}$^{el}$. Without an external pressure, \emph{V}=$\varphi$$_{V}$$\,$\emph{T}$\,$\emph{V}$_{0}$+\emph{V}$_{0}$, where $\varphi$$_{V}$ is the mean volumetric thermal expansion coefficient (usually, $\varphi$$_{V}$>0) and \emph{V}$_{0}$ is the equilibrium volume at 0 K. Ignoring high order terms (see Appendix B), $\Delta$\emph{E}$^{th}$ can be expressed as
\begin{equation}
\Delta\emph{E}^{\,th}
=-2\gamma_0\,\varphi_{V}\emph{T}\,\emph{V}_{0}(\emph{host})\,\delta\emph{V}_{q\rightarrow q'}.   \tag{(3)}
\end{equation}
\noindent Here, $\gamma$$_0$ is the elastic constant and $\delta$\emph{V}$_{q\rightarrow q'}$=\emph{V}$_{0}$($\alpha$,\emph{q'})-\emph{V}$_{0}$($\alpha$,\emph{q}) is the local volume change induced by defect $\alpha$ during the ionization from \emph{q} to \emph{q'}.

Moving to \emph{F}$^{ph}$, it can be expressed as \emph{F}$^{ph}$= $\sum_{i}$[$\frac{1}{2}$$\hbar$$\omega$$_{i}$+\emph{k}$_{B}$\emph{T}$\,$\emph{ln}$\lbrace$1-\emph{exp}(-$\frac{\hbar \omega _i}{\emph{k}_B\emph{T}}$)$\rbrace$] under the QHA \cite{32}, where $\hbar$, $\omega$$_{i}$, and \emph{k}$_{B}$ are the reduced Planck constant, phonon eigenfrequency, and Boltzmann constant, respectively. Consequently, under a first-order approximation (see Appendix C), $\Delta$\emph{F}$^{ph}$ can be written as
\begin{equation}
\Delta\emph{F}^{\,ph} =\Delta\emph{F}^{\,zp}+\Delta\widetilde{F}^{ph}
=\sum_{i}\frac{1}{2}\hbar\Delta\omega_{i}+\sum_{i}\emph{k}_{B}\emph{T}\frac{\Delta \omega _i}{\omega_i(\alpha,\emph{q})},   \tag{(4)}
\end{equation}
\noindent where $\Delta$$\omega$$_{i}$= $\omega$$_{i}$($\alpha$,\emph{q'})-$\omega$$_{i}$($\alpha$,\emph{q}) is the \emph{i}-th phonon eigenfrequency difference for defect $\alpha$ during the ionization from \emph{q} to \emph{q'}. $\Delta$\emph{F}$^{zp}$ is the contribution of zero-point vibrations and the $\Delta$$\widetilde{F}$$^{ph}$ is the pure temperature-dependent part.

\subsection*{B. Fundamental Rules}
First, we consider the role of $\Delta$\emph{F}$^{el}$ (dominated by $\Delta$\emph{E}$^{th}$) on $\varepsilon$$_{\alpha}$(\emph{q}/\emph{q'}). In a common semiconductor, rising temperature leads to volume expansion ($\varphi$$_{V}$>0). During the ionization of an acceptor (donor) from \emph{q} to \emph{q'}, the $\delta$\emph{V}$_{q\rightarrow q'}$ of the defect may expand (shrink) due to the larger (smaller) electron occupation, giving rise to a positive (negative) $\delta$\emph{V}$_{q\rightarrow q'}$. According to Eq. (3), $\Delta$\emph{E}$^{th}$ is negative (positive) and decreases (increases) with increasing temperature for an acceptor (donor), shallowing (deepening) $\varepsilon$$_{\alpha}$(\emph{q}/\emph{q'}).

Second, we consider the role of $\Delta$\emph{F}$^{ph}$ on $\varepsilon$$_{\alpha}$(\emph{q}/\emph{q'}). According to Eq. (4), the sign of $\Delta$\emph{F}$^{ph}$ is mostly determined by $\Delta$$\omega$$_{i}$. The phonon frequencies can be approximately understood using a one-dimensional harmonic oscillator model with $\omega$$\sim$$\sqrt{\frac{k}{m}}$, where the \emph{k} is the force constant for the system, capturing to the strength of atomic bonds. During the ionization of an acceptor (donor), the extra electrons are added to (removed from) the low (high) energy bonding (anti-bonding) states, which consequently stabilize the chemical bonds and enhance the bond strength surrounding the acceptor (donor). Therefore, $\Delta$$\omega$$_{i}$ is positive for both donors and acceptors. Consequently, the $\Delta$\emph{F}$^{ph}$ is positive and increases with rising temperature, deepening $\varepsilon$$_{\alpha}$(\emph{q}/\emph{q'}) for both donors and acceptors. Moreover, it is expected that $\delta$\emph{V}$_{q\rightarrow q'}$ and $\bar{m}$ (defined as the average atomic mass of the defect and its nearest-neighbor atoms) may be key factors in determining the exact value of $\Delta$\emph{F}$^{ph}$. Specifically, a larger $\delta$\emph{V}$_{q\rightarrow q'}$ indicates a larger bonding strength change around the defect during the charge-state transition, leading to the larger $\Delta$$\omega$$_{i}$ (and hence larger $\Delta$\emph{F}$^{ph}$); the larger the $\bar{m}$, the smaller $\Delta$\emph{F}$^{zp}$.

Based on the above understanding of $\Delta$\emph{F}$^{el}$ and $\Delta$\emph{F}$^{ph}$, we can propose two fundamental rules for the temperature dependence of $\varepsilon$$^{a}$$_{\alpha}$(\emph{q}/\emph{q'}) and $\varepsilon$$^{r}$$_{\alpha}$(\emph{q}/\emph{q'}), respectively. For donors, both $\Delta$\emph{F}$^{el}$ and $\Delta$\emph{F}$^{ph}$ can downshift the $\varepsilon$$^{a}$$_{\alpha}$(\emph{q}/\emph{q'}) levels towards lower energy values, and the downshift grows with temperature. Meanwhile, as shown in Fig. 1, the larger the |$\delta$\emph{V}$_{q\rightarrow q'}$| of a donor, the larger the $\Delta$\emph{E}$^{th}$ and $\Delta$\emph{F}$^{ph}$, and consequently the larger the downshift of $\varepsilon$$^{a}$$_{\alpha}$(\emph{q}/\emph{q'}). For acceptors, the (negative) $\Delta$\emph{F}$^{el}$ and (positive) $\Delta$\emph{F}$^{ph}$ have a cancelling effect, because they cause the $\varepsilon$$^{a}$$_{\alpha}$(\emph{q}/\emph{q'}) level to shift in opposite directions. Comparing Eqs. (3) and (4), it is expected that the changes of |$\Delta$\emph{E}$^{th}$| could be more significant than those of |$\Delta$$\widetilde{F}$$^{ph}$| for the variable |$\delta$\emph{V}$_{q\rightarrow q'}$|. Accordingly, as shown in Fig. 1, for an acceptor with small (large) |$\delta$\emph{V}$_{q\rightarrow q'}$|, |$\Delta$$\widetilde{F}$$^{ph}$|>|$\Delta$\emph{E}$^{th}$| (|$\Delta$$\widetilde{F}$$^{ph}$|<|$\Delta$\emph{E}$^{th}$|), which may upshift (downshift) $\varepsilon$$^{a}$$_{\alpha}$(\emph{q}/\emph{q'}) in energy. Therefore, we can propose Rule I on the changes of $\varepsilon$$^{a}$$_{\alpha}$(\emph{q}/\emph{q'}) [$\Delta$$\varepsilon$$^{a}$$_{\alpha}$(\emph{q}/\emph{q'})] in semiconductors at different temperatures. \underline{\emph{Rule I(a) for donors}}: the higher the temperature, the larger the $\Delta$$\varepsilon$$^{a}$$_{\alpha}$(\emph{q}/\emph{q'}) towards deeper levels; the larger the |$\delta$\emph{V}$_{q\rightarrow q'}$|, the larger the $\Delta$$\varepsilon$$^{a}$$_{\alpha}$(\emph{q}/\emph{q'}) towards deeper levels. \underline{\emph{Rule I(b) for acceptors}}: for the acceptors with large (small) |$\delta$\emph{V}$_{q\rightarrow q'}$|, the higher the temperature, the larger the $\Delta$$\varepsilon$$^{a}$$_{\alpha}$(\emph{q}/\emph{q'}) towards shallower (deeper) levels; the larger (smaller) the |$\delta$\emph{V}$_{q\rightarrow q'}$|, the larger the $\Delta$$\varepsilon$$^{a}$$_{\alpha}$(\emph{q}/\emph{q'}) towards shallower (deeper) levels.

\begin{figure}[htb!]
\includegraphics[width=1.0\textwidth]{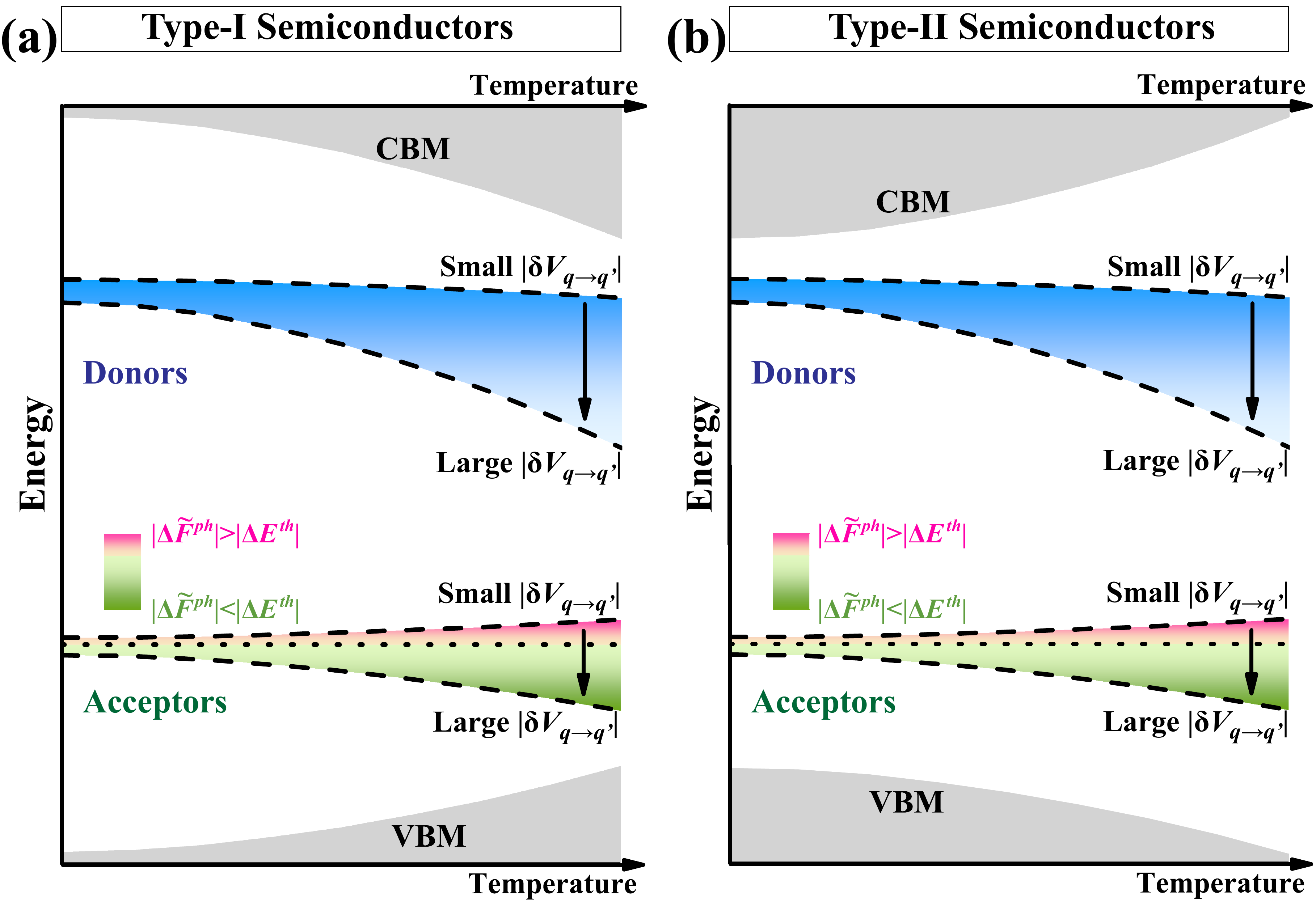}
\caption{\label{fig:structure} \textbf{Fundamental rules for temperature-dependence of $\varepsilon$$_{\alpha}$(\emph{q}/\emph{q'}).} Schematic illustration of the effects of temperature on $\varepsilon$$_{\alpha}$(\emph{q}/\emph{q'}) levels in (a) type-I and (b) type-II semiconductors, in which |$\delta$\emph{V}$_{q\rightarrow q'}$| is discovered to play a critical role. For convenience of plotting, we assume that donors (or acceptors) have similar $\varepsilon$$_{\alpha}$(\emph{q}/\emph{q'}) at 0 K [Note that $\Delta$$\varepsilon$$_{\alpha}$(\emph{q}/\emph{q'}) is independent of the initial $\varepsilon$$_{\alpha}$(\emph{q}/\emph{q'}) at 0 K]. See text for more details.}
\label{1}
\end{figure}

After having established the role of the free energy corrections, we next consider the changes in the band edge. Generally, there are two typical types of temperature-dependent band edge changes, as demonstrated in Fig. 1. In many conventional semiconductors, \emph{e.g.}, GaN \cite{33,34} and GaAs \cite{35}, the CBM (VBM) energy positions usually downshift (upshift) as temperature increases, \emph{e.g.}, $\Delta$$\varepsilon$$_{CBM}$<0 and $\Delta$$\varepsilon$$_{VBM}$>0, denoted as type-I semiconductors (Fig. 1a). Type-II semiconductors (Fig.1b), \emph{e.g.}, CsPbI$_3$ \cite{36} and MAPbI$_3$ \cite{37,38}, are opposite to the type-I cases, \emph{e.g.}, $\Delta$$\varepsilon$$_{CBM}$>0 and $\Delta$$\varepsilon$$_{VBM}$<0. Combining Rule I and specific band edge changes, we arrive at Rule II on the temperature dependence of $\varepsilon$$^{r}$$_{\alpha}$(\emph{q}/\emph{q'}) in semiconductors. \underline{\emph{Rule II(a) for donors}}: the $\varepsilon$$^{r}$$_{\alpha}$(\emph{q}/\emph{q'}) in type-I semiconductors can become shallower, deeper or stay unchanged under different temperatures (Fig. 1a), depending on the different strengths of the opposing effect between $\Delta$$\varepsilon$$^{a}$$_{\alpha}$(\emph{q}/\emph{q'}) and $\Delta$$\varepsilon$$_{CBM}$; the $\varepsilon$$^{r}$$_{\alpha}$(\emph{q}/\emph{q'}) in type-II semiconductors will always become deeper (Fig. 1b), due to the synergistic effect between $\Delta$$\varepsilon$$^{a}$$_{\alpha}$(\emph{q}/\emph{q'}) and $\Delta$$\varepsilon$$_{CBM}$. \underline{\emph{Rule II(b) for acceptors}}: the $\varepsilon$$^{r}$$_{\alpha}$(\emph{q}/\emph{q'}) with small (large) |$\delta$\emph{V}$_{q\rightarrow q'}$| in type-I (type-II) semiconductors can become either shallower, deeper or stay unchanged as a function of temperature, originating from the opposing effect between $\Delta$$\varepsilon$$^{a}$$_{\alpha}$(\emph{q}/\emph{q'}) and $\Delta$$\varepsilon$$_{VBM}$; the $\varepsilon$$^{r}$$_{\alpha}$(\emph{q}/\emph{q'}) with large (small) |$\delta$\emph{V}$_{q\rightarrow q'}$| in type-I (type-II) semiconductors will always become shallower (deeper), due to the synergistic effect between $\Delta$$\varepsilon$$^{a}$$_{\alpha}$(\emph{q}/\emph{q'}) and $\Delta$$\varepsilon$$_{VBM}$, as shown in Fig. 1a (Fig. 1b).

\subsection*{C. Verification in GaN}
Taking GaN as a prototype example, we have systematically studied the effects of temperature on $\varepsilon$$_{\alpha}$(\emph{q}/\emph{q'}) for ten different defects [see S.I in Supplementary Material (SM)]. The donor-like defects include N vacancy (V$_N$), substitutional Si$_{Ga}$, Ge$_{Ga}$, and O$_N$, while the acceptor-like defects include Mg$_{Ga}$, Zn$_{Ga}$, Be$_{Ga}$, Ca$_{Ga}$, Cd$_{Ga}$, and C$_N$ \cite{39,40}. Many of them are commonly observed in GaN \cite{39,40}.

First, we test the relationship between $\Delta$\emph{F}$^{el}$ and $\delta$\emph{V}$_{q\rightarrow q'}$ in GaN. As shown in Fig. 2a, the calculated $\Delta$\emph{F}$^{el}$ and $\Delta$\emph{E}$^{th}$ for these defects are almost identical, except for Si$_{Ga}$ and O$_{N}$ between 0 and +1 charge-state transitions at \emph{T}>800 K, confirming that the contributions of \emph{E}$^{el}$ and \emph{S}$^{el}$ to \emph{F}$^{el}$ are usually small in semiconductors \cite{12,31}. Deviations at high temperatures partially originate from the shallow-level-induced electron-occupation changes. \emph{E}$^{th}$ can be directly evaluated using first-principles calculations under hydrostatic-stress conditions \cite{42,43}, adopting the experimental $\varphi$$_{V}$ \cite{44}. As shown in Fig. 2a, the calculated $\Delta$\emph{E}$^{th}$ of these defects increase almost linearly as temperature increases. Interestingly, the calculated $\delta$\emph{V}$_{q\rightarrow q'}$ are all positive (negative) for the acceptors (donors) during ionization, resulting in the negative (positive) $\Delta$\emph{E}$^{th}$. Taking 800 K as a typical temperature, as shown in Fig. 2b, we have plotted the relationship between $\Delta$\emph{E}$^{th}$ and $\delta$\emph{V}$_{q\rightarrow q'}$ for these defects, which exhibits an almost linear dependence, confirming our expectation from Eq. (3). A similar linear dependence behavior with different slopes is observed at other temperatures [Fig.S1 in S.II].

\begin{figure}[htb!]
\includegraphics[width=1.0\textwidth]{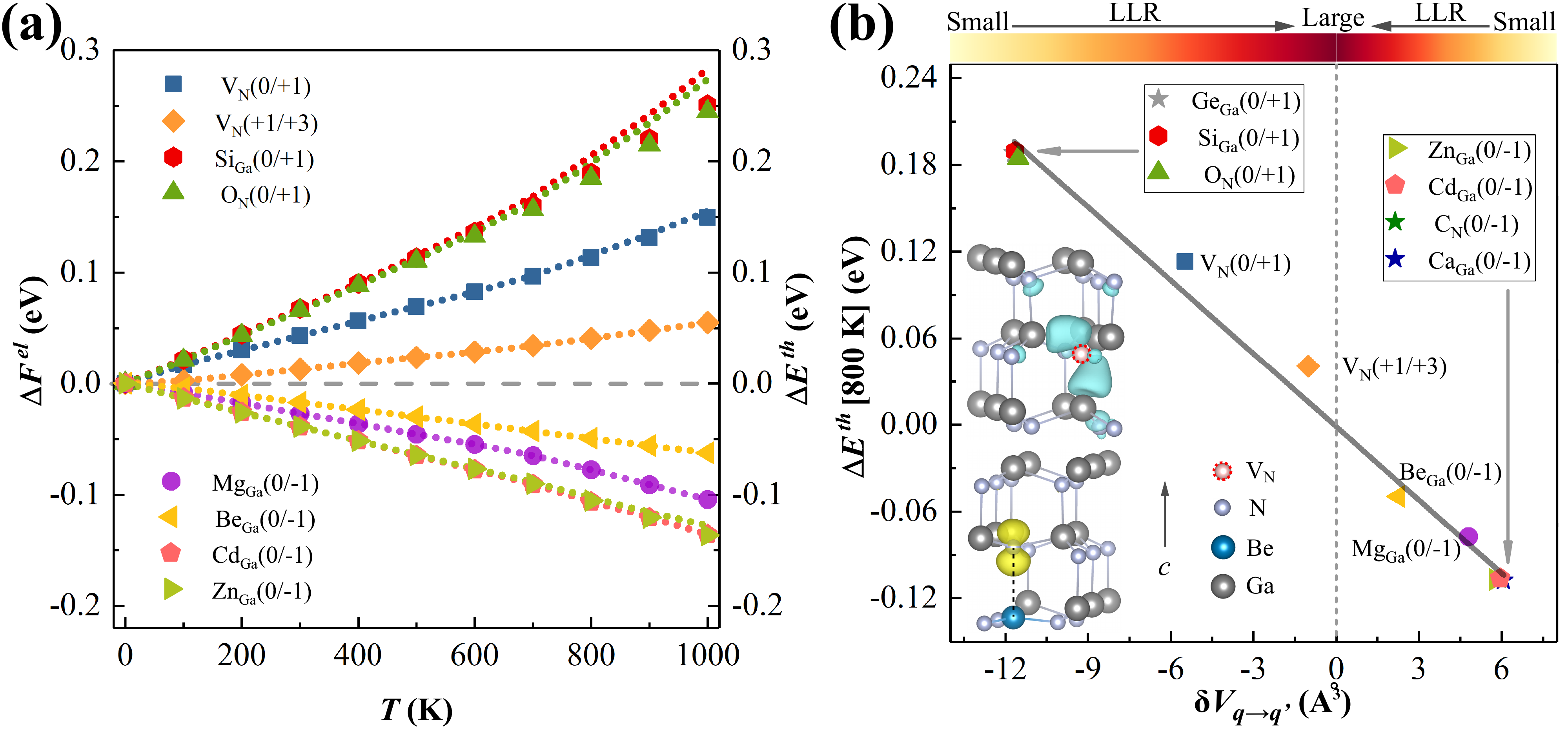}
\caption{\label{fig:structure} \textbf{Electronic contribution to the temperature dependence of $\varepsilon$$_{\alpha}$(\emph{q}/\emph{q'}) in GaN.}
(a) $\Delta$\emph{F}$^{el}$ (dashed lines) and $\Delta$\emph{E}$^{th}$ (symbols) as a function of temperature for different defects in GaN. (b) Relationship between $\delta$\emph{V}$_{q\rightarrow q'}$ and $\Delta$\emph{E}$^{th}$ for different defects at 800 K. Up-inset and bottom-inset are partial charge densities for neutral V$_N$ and Be$_{Ga}$, respectively. Black-dashed line in the bottom-inset shows the broken bond between Be and its neighboring N.}
\label{2}
\end{figure}

It is interesting to understand the origin of the diverse $\delta$\emph{V}$_{q\rightarrow q'}$ values for different defects. Overall, we discover that while $\delta$\emph{V}$_{q\rightarrow q'}$ is mainly determined by the change of LEO, the LLR can effectively compensate the LEO-induced |$\delta$\emph{V}$_{q\rightarrow q'}$|; the larger the LLR, the smaller the |$\delta$\emph{V}$_{q\rightarrow q'}$|. For example, as shown in Fig. 2b, all donors have a similar $\delta$\emph{V}$_{0\rightarrow +1}$$\sim$-12 $\mathring{A}$$^3$ except V$_N$. The $\delta$\emph{V}$_{q\rightarrow q'}$ of V$_N$ are \emph{q}/\emph{q'} dependent, \emph{e.g.}, $\delta$\emph{V}$_{0\rightarrow +1}$$\sim$-6 $\mathring{A}$$^3$ and $\delta$\emph{V}$_{+1\rightarrow +3}$$\sim$-1 $\mathring{A}$$^3$. Meanwhile, all the acceptors have similar $\delta$\emph{V}$_{0\rightarrow -1}$ $\sim$+6 $\mathring{A}$$^3$ expect Mg$_{Ga}$ ($\sim$+4.8 $\mathring{A}$$^3$) and Be$_{Ga}$ ($\sim$+2.3 $\mathring{A}$$^3$). Overall, $\delta$\emph{V}$_{q\rightarrow q'}$ is mainly determined by the change of LEO around the defect, \emph{i.e.}, the increased (decreased) LEO always significantly increases (decreases) the local volume around a defect \cite{45}, leading to a positive (negative) $\delta$\emph{V}$_{q\rightarrow q'}$. Furthermore, the $\delta$\emph{V}$_{q\rightarrow q'}$ value also depends on the strength of LLR around the defect. For example, for Si$_{Ga}$, there is a negligible LLR during the ionization (Fig. S2 in S.II), therefore, the large $\delta$\emph{V}$_{0\rightarrow +1}$ $\sim$-12 $\mathring{A}$$^3$ is mainly induced by the decreased LEO around Si$_{Ga}$. The cases of Ge$_{Ga}$ and O$_{N}$ are similar to that of Si$_{Ga}$, resulting in similar $\delta$\emph{V}$_{0\rightarrow +1}$ values. For V$_N$, due to the broken ionic bonds around V$_N$, the extra electrons from the dangling bonds (DBs) are strongly localized around V$_N$ (up-inset, Fig. 2b). During the ionization from 0 to +1, the DB electrons could be partially compensated, which consequently reduces electron screening and enhances Coulomb repulsion between the neighboring Ga$^{+3}$ ions around V$_N$. As a result, the large LLR effect around V$_N$ (Fig. S3 in S.II) effectively expands the local volume and partially compensates the initial local volume shrinkage induced by the decreased LEO. Therefore, the $\delta$\emph{V}$_{0\rightarrow +1}$ of V$_N$ is significantly smaller than that of Si$_{Ga}$. In a similar way, the $\delta$\emph{V}$_{+1\rightarrow +3}$ of V$_N$ can be further reduced from -6 to -1 $\mathring{A}$$^3$, due to the further enhanced LLR effect (Fig. S3 in S.II).

Similar to Si$_{Ga}$, a negligible LLR also exists in acceptors such as Zn$_{Ga}$, C$_N$, Ca$_{Ga}$ and Cd$_{Ga}$ (Fig. S4 in S.II). As a result, the increased LEO gives rise to a large $\delta$\emph{V}$_{0\rightarrow -1}$ $\sim$6 $\mathring{A}$$^3$ for these acceptors. However, for Be$_{Ga}$, the smaller atomic size of Be compared to Ga induces one broken ionic-bond around Be$_{Ga}$ along the \emph{c} direction, resulting in a DB hole on the broken N bond (bottom-inset, Fig. 2b) \cite{46}. During the ionization, the DB hole is fully compensated, resulting in a strongly enhanced Coulomb attraction that restores the Be-N bond along the \emph{c} direction and shrinks the local volume around Be$_{Ga}$ (Fig. S5 in S.II). This large local volume shrinkage induced by the LLR effect largely compensate the initial local volume expansion induced by the increased LEO. Hence, compared to Cd$_{Ga}$, the $\delta$\emph{V}$_{0\rightarrow -1}$ of Be$_{Ga}$ is reduced to +2.3 $\mathring{A}$$^3$. The strength of LLR in Mg$_{Ga}$ (Fig. S6 in S.II) is between Be$_{Ga}$ and Cd$_{Ga}$, resulting in an intermediate $\delta$\emph{V}$_{0\rightarrow -1}$ value between that of Be$_{Ga}$ and Cd$_{Ga}$. Therefore, as shown in Fig. 2b, we conclude that the variable $\delta$\emph{V}$_{q\rightarrow q'}$ in different defects are mainly determined by the competing effect between LEO and LLR.

Second, we explore the relationship between $\Delta$\emph{F}$^{ph}$ and $\delta$\emph{V}$_{q\rightarrow q'}$ in GaN. As shown in Fig. 3a, the $\Delta$\emph{F}$^{ph}$ of both donors and acceptors are positive and increase with increasing temperature. Here, we focus on the $\Delta$$\widetilde{F}$$^{ph}$, which determines the temperature dependence of $\Delta$\emph{F}$^{ph}$. We expect that |$\delta$\emph{V}$_{q\rightarrow q'}$| and $\bar{m}$ are the two main factors in determining $\Delta$$\widetilde{F}$$^{ph}$. Indeed, the $\Delta$$\widetilde{F}$$^{ph}$ values of all the defects at different temperatures can be well fitted by a simple but unified formula given as
\begin{equation}
\Delta\widetilde{F}^{ph}=\emph{k}_{B}\emph{T}\,(\emph{a}_{1}|\delta\emph{V}_{q\rightarrow q'}|+\emph{a}_{2}|\delta\emph{V}_{q\rightarrow q'}|^2
+\emph{b}_{1}\bar{m}+\emph{b}_{2}\bar{m}^2+\emph{c}|\delta\emph{V}_{q\rightarrow q'}|\bar{m}+\emph{d}).    \tag{(5)}
\end{equation}
Fig. 3b shows the case of \emph{T}=800 K. Similar behaviors are observed at other temperatures but with different parameter values (Fig.S7 in S.II). Overall, it is found that the |$\delta$\emph{V}$_{q\rightarrow q'}$| term is the dominant factor for $\Delta$$\widetilde{F}$$^{ph}$, and a larger |$\delta$\emph{V}$_{q\rightarrow q'}$| usually gives a larger $\Delta$$\widetilde{F}$$^{ph}$. For example, the larger $\Delta$$\widetilde{F}$$^{ph}$ of Si$_{Ga}$ compared to Cd$_{Ga}$ (116 \emph{v.s.} 44 meV) is mainly due to its larger |$\delta$\emph{V}$_{q\rightarrow q'}$| ($\sim$12 \emph{v.s.} $\sim$6 $\mathring{A}$$^3$). For defects with similar |$\delta$\emph{V}$_{q\rightarrow q'}$| values, $\bar{m}$ becomes important in determining $\Delta$$\widetilde{F}$$^{ph}$, and a smaller $\bar{m}$ gives a larger $\Delta$$\widetilde{F}$$^{ph}$. For example, comparing Ca$_{Ga}$, Zn$_{Ga}$, and Cd$_{Ga}$, which have a similar |$\delta$\emph{V}$_{q\rightarrow q'}$|$\sim$6 $\mathring{A}$$^3$, Ca$_{Ga}$ with smaller $\bar{m}$ (19.2) than Zn$_{Ga}$ (24.2) and Cd$_{Ga}$ (33.6) has larger $\Delta$$\widetilde{F}$$^{ph}$ (80 meV) compared to Zn$_{Ga}$ (47 meV) and Cd$_{Ga}$ (44 meV). We notice that the calculated $\Delta$$\widetilde{F}$$^{ph}$ value for C$_{N}$ at 600 K also agrees with a previous study \cite{19}.

\begin{figure}[htb!]
\includegraphics[width=1.0\textwidth]{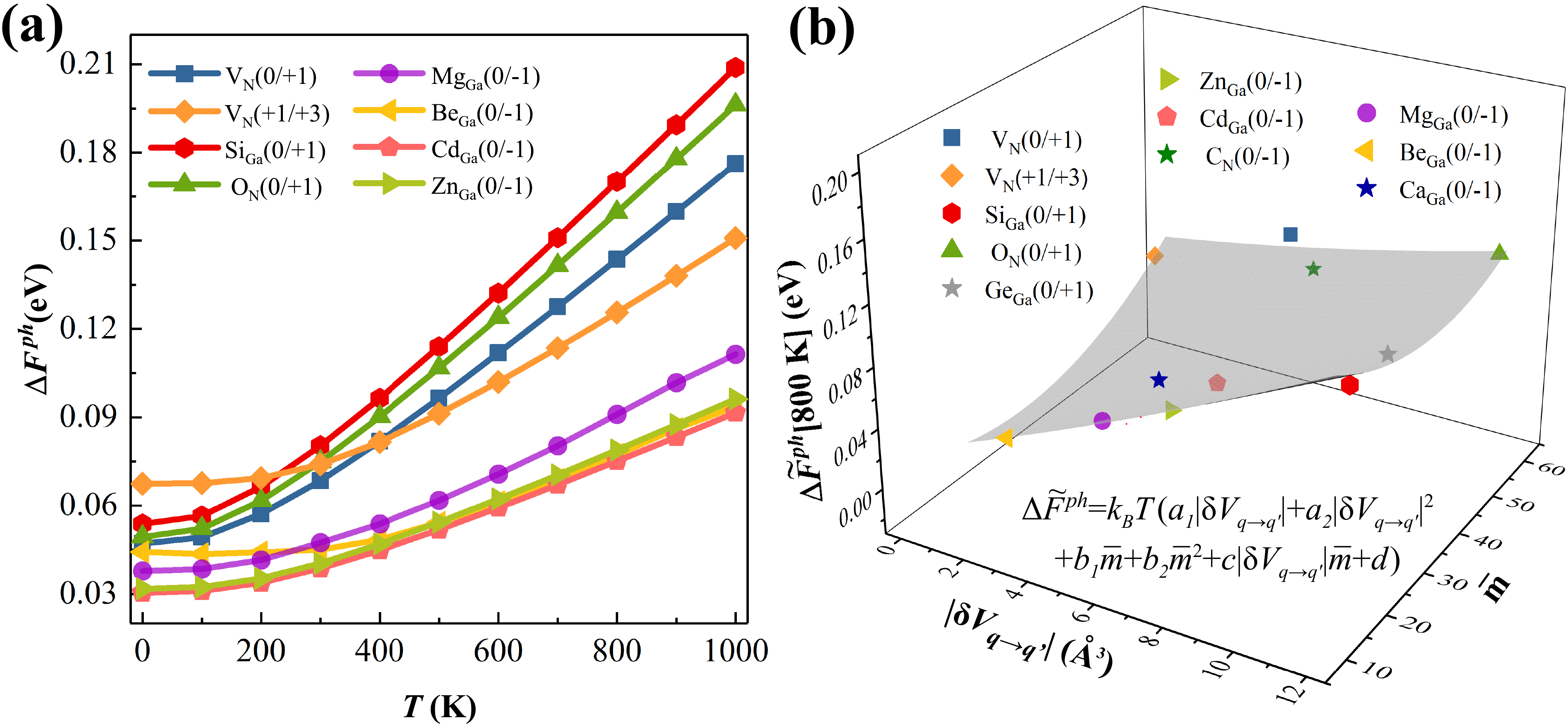}
\caption{\label{fig:structure} \textbf{Vibrational contribution to the temperature dependence of $\varepsilon$$_{\alpha}$(\emph{q}/\emph{q'}) in GaN.}
(a) $\Delta$\emph{F}$^{ph}$ as a function of temperature for different defects in GaN. (b) Relationship between $\Delta$$\widetilde{F}$$^{ph}$, |$\delta$\emph{V}$_{q\rightarrow q'}$| and $\bar{m}$, which can be described by a general polynomial formula. At 800 K, \emph{a}$_{1}$=0.15, \emph{a}$_{2}$=0.0017, \emph{b}$_{1}$=-0.0263, \emph{b}$_{2}$=0.0006, \emph{c}=-0.0019, and \emph{d}=0.4311.}
\label{3}
\end{figure}

Combining the results of $\Delta$\emph{F}$^{el}$ (Fig. 2a) and $\Delta$\emph{F}$^{ph}$ (Fig. 3a), we arrive at the temperature-dependent $\varepsilon$$_{\alpha}$(\emph{q}/\emph{q'}) of defects in GaN. GaN is a type-I semiconductor \cite{33}, whose |$\Delta$$\varepsilon$$_{CBM}$| (Fig. 4a) is noticeably larger than |$\Delta$$\varepsilon$$_{VBM}$| (Fig. 4b) at a given temperature, due to the different band-edge orbital characters \cite{33,49}. Importantly, the calculated bandgap of GaN as a function of temperature agrees well with the experimental measurements \cite{34}, confirming the reliability of our computational methods (Fig. S8 in S.II).

As shown in Fig. 4a, three typical donors, \emph{i.e.}, V$_N$, Si$_{Ga}$, and O$_N$, are selected to demonstrate the temperature dependence of $\varepsilon$$_{\alpha}$(\emph{q}/\emph{q'}) for donors, to verify our proposed Rules I(a) and II(a). Interestingly, these donors exhibit quite different temperature-dependent behaviors. Without the consideration of $\Delta$$\varepsilon$$_{VBM}$, the $\varepsilon$$^{a}$$_{\alpha}$(\emph{q}/\emph{q'}) of all the donors become deeper as the temperature increases, \emph{i.e.}, the higher the temperature, the deeper the $\varepsilon$$^{a}$$_{\alpha}$(\emph{q}/\emph{q'}). Surprisingly, the $\varepsilon$$^{a}$$_{\alpha}$(0/+1) of Si$_{Ga}$ and V$_N$, which have similar shallow levels at \emph{T}=0 K (with the inclusion of $\Delta$\emph{F}$^{zp}$ contribution), exhibit dramatically different temperature dependences, \emph{i.e.}, the change of $\Delta$$\varepsilon$$^{a}$$_{\alpha}$(0/+1) in Si$_{Ga}$ (-0.44 eV) in the range 0<\emph{T}<1000 K is much larger than that in V$_N$ (-0.28 eV), due to the significantly larger |$\Delta$\emph{V}$_{0\rightarrow +1}$| in Si$_{Ga}$ (Fig. 2b). Interestingly, although the $\varepsilon$$^{a}$$_{\alpha}$(0/+1) of O$_N$ is much deeper than that of Si$_{Ga}$ at 0 K, they exhibit almost the same trend of $\Delta$$\varepsilon$$^{a}$$_{\alpha}$(0/+1) under different temperatures, due to their similar |$\Delta$\emph{V}$_{0\rightarrow +1}$| (the slight difference at high temperatures is caused by their different $\bar{m}$). Unexpectedly, the $\Delta$$\varepsilon$$^{a}$$_{\alpha}$(\emph{q}/\emph{q'}) of one defect can also exhibit totally different behaviors under different charge-state transitions. For example, the $\Delta$$\varepsilon$$^{a}$$_{\alpha}$(0/+1) and $\Delta$$\varepsilon$$^{a}$$_{\alpha}$(+1/+3) for V$_N$ are dramatically different because of their largely different |$\delta$\emph{V}$_{q\rightarrow q'}$| (Fig. 2b). The above observations, along with other calculated donors (Fig. S9a in S.II), confirm the proposed Rule I(a) on the relationship between |$\delta$\emph{V}$_{q\rightarrow q'}$| and $\Delta$$\varepsilon$$^{a}$$_{\alpha}$(\emph{q}/\emph{q'}) for donors at different temperatures.

Combining $\varepsilon$$^{a}$$_{\alpha}$(\emph{q}/\emph{q'}) with the calculated CBM bowing of GaN, we obtain the $\varepsilon$$^{r}$$_{\alpha}$(\emph{q}/\emph{q'}) of donors. Interestingly, as exhibited in Fig. 4c, the $\varepsilon$$^{r}$$_{\alpha}$(\emph{q}/\emph{q'}) can become either shallower [$\varepsilon$$^{r}$$_{\alpha}$(\emph{q}/\emph{q'})>0], deeper [$\varepsilon$$^{r}$$_{\alpha}$(\emph{q}/\emph{q'})<0] or even unchanged [$\varepsilon$$^{r}$$_{\alpha}$(\emph{q}/\emph{q'})$\sim$0] for different donors in different temperature regions. For examples, for Si$_{Ga}$ and O$_N$, $\Delta$$\varepsilon$$^{r}$$_{\alpha}$(0/+1)$\approx$0 in the range 0<\emph{T}<500 K, due to the largest opposing effect between |$\Delta$$\varepsilon$$_{CBM}$| and |$\Delta$$\varepsilon$$^{a}$$_{\alpha}$(0/+1)| (|$\Delta$$\varepsilon$$_{CBM}$|$\approx$|$\Delta$$\varepsilon$$^{a}$$_{\alpha}$(0/+1)|); for \emph{T}>500 K, |$\Delta$$\varepsilon$$_{CBM}$|<|$\Delta$$\varepsilon$$^{a}$$_{\alpha}$(0/+1)| gives rise to $\Delta$$\varepsilon$$^{r}$$_{\alpha}$(0/+1)<0, \emph{e.g.}, $\Delta$$\varepsilon$$^{r}$$_{\alpha}$(0/+1) of Si$_{Ga}$ is -0.04 eV at \emph{T}=1000 K. For V$_N$, |$\Delta$$\varepsilon$$_{CBM}$|> |$\Delta$$\varepsilon$$^{a}$$_{\alpha}$(0/+1)| in the range 0<\emph{T}<1000 K, resulting in $\Delta$$\varepsilon$$^{r}$$_{\alpha}$(\emph{q}/\emph{q'})>0. Among all the $\Delta$$\varepsilon$$^{r}$$_{\alpha}$(\emph{q}/\emph{q'}), the largest value occurs in the $\Delta$$\varepsilon$$^{r}$$_{\alpha}$(+1/+3) of V$_N$ (0.33 eV at \emph{T}=1000 K), due to the smallest opposing effect between |$\Delta$$\varepsilon$$_{CBM}$| and |$\Delta$$\varepsilon$$^{a}$$_{\alpha}$(+1/+3)|. The above observations, along with other calculated donors (Fig. S9a in S.II), confirm our proposed Rule II(a) that the $\Delta$$\varepsilon$$^{r}$$_{\alpha}$(\emph{q}/\emph{q'}) of donors in type-I semiconductors is determined by the relative magnitude and sign of $\Delta$$\varepsilon$$^{a}$$_{\alpha}$(\emph{q}/\emph{q'}) and $\Delta$$\varepsilon$$_{CBM}$.

\begin{figure}[htb!]
\includegraphics[width=1.0\textwidth]{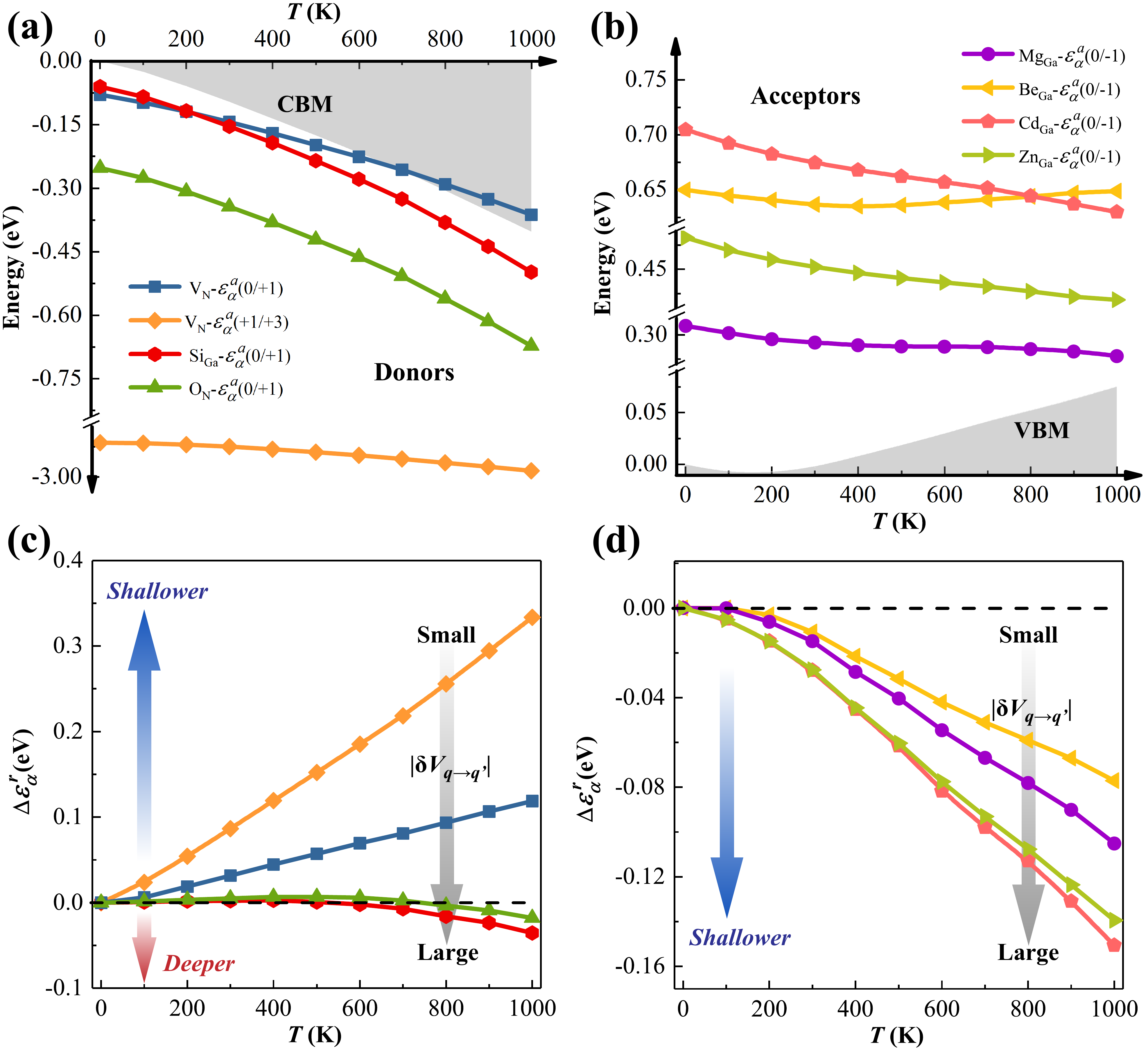}
\caption{\label{fig:structure} \textbf{Temperature dependence of $\varepsilon$$_{\alpha}$(\emph{q}/\emph{q'}) in GaN.}
$\varepsilon$$^{a}$$_{\alpha}$(\emph{q}/\emph{q'}) levels of several typical (a) donors and (b) acceptors in GaN as a function of temperature. Calculated temperature-dependent CBM and VBM changes are also plotted in (a) and (b), respectively. (c) and (d) are corresponding $\Delta$$\varepsilon$$^{r}$$_{\alpha}$(\emph{q}/\emph{q'}) for (a) and (b), respectively. Gray arrow indicates the trend of the sizes of |$\delta$\emph{V}$_{q\rightarrow q'}$|.}
\label{4}
\end{figure}

In Fig. 4b, four typical acceptors, \emph{i.e.}, Zn$_{Ga}$, Mg$_{Ga}$, Be$_{Ga}$, and Cd$_{Ga}$, are selected to demonstrate the temperature dependence of $\varepsilon$$_{\alpha}$(\emph{q}/\emph{q'}) for acceptors, to verify our proposed Rules I(b) and II(b). Holding large |$\Delta$\emph{V}$_{0\rightarrow -1}$|, the $\varepsilon$$^{a}$$_{\alpha}$(0/-1) of Cd$_{Ga}$ and Zn$_{Ga}$ always becomes shallower [\emph{i.e.}, $\varepsilon$$^{a}$$_{\alpha}$(0/-1)<0] as temperature increases. Interestingly, regardless of the significantly different $\varepsilon$$^{a}$$_{\alpha}$(0/-1) values at 0 K for Cd$_{Ga}$ and Zn$_{Ga}$, their $\Delta$$\varepsilon$$^{a}$$_{\alpha}$(0/-1) exhibit similar temperature dependences, mostly due to their similar |$\Delta$\emph{V}$_{0\rightarrow -1}$| (Fig. 2b). Again, their slightly different $\Delta$$\varepsilon$$^{a}$$_{\alpha}$(0/-1) at high temperatures could be due to their different $\bar{m}$. Surprisingly, although the Cd$_{Ga}$ and Be$_{Ga}$ have close $\varepsilon$$^{a}$$_{\alpha}$(0/-1) values at 0 K, their $\Delta$$\varepsilon$$^{a}$$_{\alpha}$(0/-1) exhibit different (even opposite) temperature dependences, due to their significantly different |$\Delta$\emph{V}$_{0\rightarrow -1}$| (Fig. 2b); with small |$\Delta$\emph{V}$_{0\rightarrow -1}$|, the $\varepsilon$$^{a}$$_{\alpha}$(0/-1) of Be$_{Ga}$ becomes even deeper when \emph{T}>400 K, swapping the relative positions of Cd$_{Ga}$ and Be$_{Ga}$ at \emph{T}=0 K and \emph{T}=1000 K. Since the |$\Delta$\emph{V}$_{0\rightarrow -1}$| of Mg$_{Ga}$ is in between Cd$_{Ga}$ and Be$_{Ga}$, the $\Delta$$\varepsilon$$^{a}$$_{\alpha}$(0/-1) of Mg$_{Ga}$ is $\sim$0 in the range 300<\emph{T}<700 K. These observations, along with other calculated acceptors (Fig. S9b in S.II), confirm the Rule I(b) on the $\Delta$$\varepsilon$$^{a}$$_{\alpha}$(\emph{q}/\emph{q'}) of acceptors, especially the critical role played by |$\delta$\emph{V}$_{q\rightarrow q'}$|. We emphasize that the values of |$\Delta$$\varepsilon$$^{a}$$_{\alpha}$(\emph{q}/\emph{q'})| for acceptors are usually much smaller than those for donors, due to the large cancelling effect between $\Delta$\emph{F}$^{el}$ and $\Delta$\emph{F}$^{ph}$ for acceptors.

Combining $\varepsilon$$^{a}$$_{\alpha}$(\emph{q}/\emph{q'}) with the calculated VBM bowing of GaN, we can obtain the $\varepsilon$$^{r}$$_{\alpha}$(\emph{q}/\emph{q'}) of acceptors. Overall, as shown in Fig. 4d, the $\varepsilon$$^{r}$$_{\alpha}$(0/-1) of all acceptors in GaN becomes shallower [\emph{i.e.}, $\Delta$$\varepsilon$$^{r}$$_{\alpha}$(0/-1)<0]. For Cd$_{Ga}$ and Zn$_{Ga}$, the synergistic effect between $\Delta$$\varepsilon$$^{a}$$_{\alpha}$(0/-1) and $\Delta$$\varepsilon$$_{VBM}$ results in a large value of $\Delta$$\varepsilon$$^{r}$$_{\alpha}$(0/-1), \emph{e.g.}, $\Delta$$\varepsilon$$^{r}$$_{\alpha}$(0/-1)=-0.15 eV at \emph{T}=1000 K. For Mg$_{Ga}$ with $\Delta$$\varepsilon$$^{a}$$_{\alpha}$(0/-1)$\sim$0, its $\Delta$$\varepsilon$$^{r}$$_{\alpha}$(0/-1) closely follows the trend of $\Delta$$\varepsilon$$_{VBM}$. For Be$_{Ga}$, the opposing effect between $\Delta$$\varepsilon$$^{a}$$_{\alpha}$(0/-1) and $\Delta$$\varepsilon$$_{VBM}$ leads to a small $\Delta$$\varepsilon$$^{r}$$_{\alpha}$(0/-1). The above observations, along with other calculated acceptors (Fig. S9b in S.II), confirm our proposed Rule II(b) that the $\varepsilon$$^{r}$$_{\alpha}$(\emph{q}/\emph{q'}) of acceptors in type-I semiconductors, depending on the size of |$\delta$\emph{V}$_{q\rightarrow q'}$|, can exhibit either synergistic or opposing effects between $\varepsilon$$^{a}$$_{\alpha}$(\emph{q}/\emph{q'}) and $\Delta$$\varepsilon$$_{VBM}$.
\section*{III. Conclusion and Outlook}
We emphasize that, although the overall sizes of |$\Delta$$\varepsilon$$_{\alpha}$(\emph{q}/\emph{q'})| are not huge in GaN (generally <0.4 eV), we expect that the temperature effect, obeying the same rules as we have developed, could be much more noticeable in many other systems, \emph{e.g.}, superhard semiconductors (\emph{e.g.}, diamond, in which defects play a key role for realizing quantum bits) or organic-inorganic hybrid perovskites (\emph{e.g.}, MAPbI$_3$, in which defects play a key role for limiting their solar efficiencies), in which the phonon vibrations or band edge changes are much more significant than those in GaN. Since the carrier concentrations and defect-mediated non-radiative carrier recombination in semiconductors are very sensitive to the positions of $\varepsilon$$_{\alpha}$(\emph{q}/\emph{q'}) inside the bandgap that are temperature dependent, our theory may also be applied to reexamine or explain many existing puzzles on the disagreements between experimental measurements and static first-principles calculations.

In conclusion, we have derived the basic formulas and consequently established two fundamental rules for the temperature dependence of $\varepsilon$$_{\alpha}$(\emph{q}/\emph{q'}) for both donors and acceptors in semiconductors, a question that has remained unanswered for decades. As we demonstrated in GaN, the temperature-driven changes of $\varepsilon$$_{\alpha}$(\emph{q}/\emph{q'}) for different defects can be rather diverse, \emph{i.e.}, it can become shallower, deeper or stay unchanged. The ultimate behavior is mainly determined by the synergistic or opposing effects between free energy corrections and band edge changes. In particular, we discover a previously ignored physical quantity, $\delta$\emph{V}$_{q\rightarrow q'}$, that plays an unexpectedly central role in determining the temperature evolution of $\varepsilon$$_{\alpha}$(\emph{q}/\emph{q'}). Generally, these basic formulas and fundamental rules may potentially be applied to design novel semiconductor devices operated under high or varying temperatures.

\section*{Acknowledgements}
We thank Drs. J. B. Chen and P. Li for helpful discussions. S.-H.W. and B.H. acknowledge support from the NSFC (Grant Nos. 11634003, 12088101) and NSAF U1930402. S. Q. acknowledges support from the NSFC (Grant No. 12047508) and Chinese Postdoctoral Science Foundation (Grant No. 2021M690329). Y.-N.W. acknowledges support from the Program for Professor of Special Appointment (Eastern Scholar TP2019019). B.M. acknowledges support from the Gianna Angelopoulos Programme for Science, Technology, and Innovation and from the Winton Programme for the Physics of Sustainability. Part of the calculations were performed at Tianhe2-JK at Computational Science Research Center.

S.Q. and Y.-N.W. contributed equally to this work.

\section*{APPENDIX A: Derivation of $\varepsilon$$_{\alpha}$(${q}$/${q'}$) at finite temperature}
Without the inclusion of temperature effects, the formation energy of a defect $\alpha$ in charge state \emph{q} is defined as \cite{12,13,50}:
\begin{equation}
\Delta\emph{H}_{\emph{f}}(\alpha,\emph{q})=\emph{E}(\alpha,\emph{q})-\emph{E}(\emph{host})+\sum\emph{n}_i\emph{E}(\emph{i})
+\sum\emph{n}_i\mu_i+\emph{q}\varepsilon_{VBM}(\emph{host})+\emph{qE}_F. \tag{(A1)}
\end{equation}
\noindent \emph{E}($\alpha$,\emph{q}) is the total energy of the supercell with defect $\alpha$ in charge state \emph{q}, whereas \emph{E}(\emph{host}) is the total energy of perfect host without defect or impurity. \emph{E}(\emph{i}) is the energy of the elemental constituent \emph{i} at its elemental monomeric phases, and $\mu$$_i$ is its chemical potential refer to \emph{E}(\emph{i}). \emph{n}$_i$ is the number of atoms exchanged with the external environment during the formation of defects for element \emph{i}, and the charge state \emph{q} is the number of electrons transferred from the supercell to the reservoirs. $\varepsilon$$_{VBM}$(\emph{host}) is the valence band maximum (VBM) of the host material and \emph{E}$_F$ is the Fermi energy refer to $\varepsilon$$_{VBM}$(\emph{host}).

The $\varepsilon$$_{\alpha}$(\emph{q}/\emph{q'}) is the Fermi level at which the charge state \emph{q} has the same formation energy with \emph{q'}:
\begin{equation}
\varepsilon_{\alpha}(\emph{q}/\emph{q'})_{w.o.T}=\frac{E(\alpha,q')-E(\alpha,q)}{\emph{q}-\emph{q'}}-\varepsilon_{VBM}(\emph{host}).   \tag{(A2)}
\end{equation}

With the inclusion of temperature effects, the $\Delta$\emph{H}$_{\emph{f}}$(\emph{$\alpha$},\emph{q}) in Eq.(A1) is replaced by the Gibbs free energy $\Delta$\emph{G}$_{\emph{f}}$(\emph{$\alpha$},\emph{q}), and the \emph{E}($\alpha$,\emph{q}) [\emph{E}($\alpha$,\emph{q'})] is replaced by free energy \emph{F}($\alpha$,\emph{q}) [\emph{F}($\alpha$,\emph{q'})]
\begin{align}
\Delta\emph{G}_{\emph{f}}(\alpha,\emph{q})[\emph{P},\emph{T}\,]=&\emph{F}(\alpha,\emph{q})[\emph{V},\emph{T}\,]
-\emph{F}(\emph{host})[\emph{V}_{host},\emph{T}\,]+(\emph{V}-\emph{V}_{host})\emph{P}+
\sum\emph{n}_i\emph{F}(\emph{i})[\emph{V}_i,\emph{T}\,]\nonumber \\
&+\sum\emph{n}_i\mu_i[\emph{P},\emph{T}\,]+\emph{q}\varepsilon_{VBM}(\emph{host})[\emph{V}_{host},\emph{T}\,]+\emph{qE}_F[\emph{V},\emph{T}\,], \tag{(A3)}
\end{align}
\noindent where \emph{P} and \emph{T} are the pressure and temperature, respectively. \emph{V} and \emph{V}$_{host}$ are the volumes of the system with defects and the host system under pressure \emph{P}, respectively. In our case of no external pressure (\emph{P}=0), Eq.(A3) can be written as
\begin{align}
\Delta\emph{G}_{\emph{f}}(\alpha,\emph{q})[\emph{P},\emph{T}\,]=&\emph{F}(\alpha,\emph{q})[\emph{V},\emph{T}\,]
-\emph{F}(\emph{host})[\emph{V},\emph{T}\,]+\sum\emph{n}_i\emph{F}(\emph{i})[\emph{V}_i,\emph{T}\,]\nonumber \\
&+\sum\emph{n}_i\mu_i[\emph{P},\emph{T}\,]+\emph{q}\varepsilon_{VBM}(\emph{host})[\emph{V},\emph{T}\,]+\emph{qE}_F[\emph{V},\emph{T}\,]. \tag{(A4)}
\end{align}

Accordingly, with the inclusion of temperature effects, $\varepsilon$$_{\alpha}$(\emph{q}/\emph{q'}) is given by
\begin{equation}
\varepsilon_{\alpha}(\emph{q}/\emph{q'})[\emph{V},\emph{T}\,] = \frac{F(\alpha,q')[\emph{V},\emph{T}\,]-F(\alpha,q)[\emph{V},\emph{T}\,]}{\emph{q}-\emph{q'}}
-\varepsilon_{VBM}(\emph{host})[\emph{V},\emph{T}\,].  \tag{(A5)}
\end{equation}

The free energy \emph{F} can be expanded around the equilibrium position as
\begin{equation}
\emph{F}(\lbrace\textbf{R}_I\rbrace)=\emph{F}_{0}+\frac{1}{2}\sum_{k,l}\emph{u}_{k}\emph{u}_{l}[\frac{\partial^2F}{\partial R_k \partial R_l}]_{\lbrace \textbf{R}_I^{0} \rbrace}+\emph{O}(\emph{u}^{3}),  \tag{(A6)}
\end{equation}
\noindent where \emph{R}$_I$ are the atomic coordinates of atom \emph{I}, and \emph{R}$_I$$^0$ are the equilibrium position. \emph{u}$_{k}$, defined as \emph{R}$_k$-\emph{R}$_k$$^0$, are the atomic displacements of atom \emph{k} from the equilibrium positions. The first and second terms are the electron and phonon free energies, respectively.
Accordingly, \emph{F}$_{0}$ includes two parts, \emph{E} (the total energy of the system without the consideration of temperature effects) and \emph{F}$^{el}$ (the corrections of the free energy induced by the electron contribution). Ignoring high order terms, we have
\begin{equation}
\emph{F}\,[\emph{V},\emph{T}\,]=\emph{E}+\emph{F}^{\,el}[\emph{V},\emph{T}\,]+\emph{F}^{\,ph}[\emph{V},\emph{T}\,],  \tag{(A7)}
\end{equation}
\noindent where \emph{F}$^{ph}$ is the correction of the free energy induced by the phonon vibration.

Under the quasi-harmonic approximation (QHA), \emph{F}$^{el}$ can be written as \cite{13,29,30}
\begin{equation}
\emph{F}^{\,el}[\emph{V},\emph{T}\,]=\emph{E}^{\,th}[\emph{V},\emph{T}\,]+\emph{E}^{\,el}[\emph{V},\emph{T}\,]
-\emph{T}\emph{S}^{\,el}[\emph{V},\emph{T}\,],  \tag{(A8)}
\end{equation}
\noindent where the first, second, and third terms are the contributions from thermal expansion, electron-occupation change, and electronic entropy, respectively. Generally, the contributions of the \emph{E}$^{el}$ and \emph{S}$^{el}$ are negligible in semiconductors, as also verified in our calculations [Fig2(a)]. Therefore, we focus on the \emph{E}$^{th}$ term in \emph{F}$^{el}$.

Under the QHA, \emph{F}$^{ph}$ can be written as \cite{32}
\begin{equation}
\emph{F}^{\,ph}= \sum_{i}[\frac{1}{2}\hbar\omega_{i}+\emph{k}_{B}\emph{T}\,\emph{ln}\lbrace 1-\emph{exp}(-\frac{\hbar \omega _i}{\emph{k}_B\emph{T}})\rbrace],  \tag{(A9)}
\end{equation}
\noindent where the $\hbar$, $\omega$$_{i}$, and \emph{k}$_{B}$ are the reduced Planck constant, phonon eigenfrequency, and Boltzmann constant, respectively.

Combining Eq.(A6) to Eq.(A9), we can obtain the $\varepsilon$$_{\alpha}$(\emph{q}/\emph{q'}) of defects under different temperatures
\begin{equation}
\varepsilon_{\alpha}(\emph{q}/\emph{q'})[\emph{V},\emph{T}\,] = \varepsilon_{\alpha}(\emph{q}/\emph{q'})_{w.o.T} + \frac{\Delta\emph{F}^{\,el}[\emph{V},\emph{T}\,] + \Delta\emph{F}^{\,ph}[\emph{V},\emph{T}\,]}{\emph{q}-\emph{q'}} - \Delta\varepsilon_{VBM}(\emph{host})[\emph{V},\emph{T}\,].  \tag{(A10)}
\end{equation}
\noindent The second term of Eq.(A10) represents the correction on the free energy differences between the \emph{q} and \emph{q'} configurations induced by the electronic ($\Delta$\emph{F}$^{el}$) and vibrational ($\Delta$\emph{F}$^{ph}$) contributions. The third term of Eq.(A10) represents the correction on the VBM energy position induced by thermal expansion and electron-phonon coupling ($\Delta$$\varepsilon$$_{VBM}$ = $\Delta$$\varepsilon$$^{th}$$_{VBM}$+$\Delta$$\varepsilon$$^{ph}$$_{VBM}$).

Without consideration of the external pressure, volume expansion induced by rising temperature can be described by the thermal expansion coefficient. From the definition of mean volumetric thermal expansion coefficient, we can obtain that \emph{T} and \emph{V} are correlated by \cite{S24}
\begin{equation}
\emph{V}=\varphi_{V}\,\emph{T}\,\emph{V}_{0}+\emph{V}_{0},  \tag{(A11)}
\end{equation}
\noindent where $\varphi$$_{V}$ is the mean volumetric thermal expansion coefficient, and \emph{V}$_{0}$ is the equilibrium volume of the system at 0 K. Thus, in case of no external pressure, we can keep \emph{T} as the only variable in our formula, and the $\varepsilon$$_{\alpha}$(\emph{q}/\emph{q'}) at a finite temperature becomes
\begin{equation}
\varepsilon_{\alpha}(\emph{q}/\emph{q'})[\emph{T}\,] = \varepsilon_{\alpha}(\emph{q}/\emph{q'})_{w.o.T} + \frac{\Delta\emph{F}^{\,el}[\emph{T}\,] + \Delta\emph{F}^{\,ph}[\emph{T}\,]}{\emph{q}-\emph{q'}} - \Delta\varepsilon_{VBM}(\emph{host})[\emph{T}\,].   \tag{(A12)}
\end{equation}

At a given temperature , $\Delta$\emph{F}$^{el}$($\Delta$\emph{E}$^{th}$) and $\Delta$$\varepsilon$$^{th}$$_{VBM}$ can be directly calculated via first-principles calculations under hydrostatic-stress conditions \cite{42,43} (it is noted that the lattice constant after thermal expansion is determined by the experiment thermal expansion coefficients of GaN \cite{S9}), $\Delta$\emph{F}$^{ph}$ can be determined through Eq.(A9) and the calculations of phonon eigenfrequencies for charge states \emph{q} and \emph{q'}. Finally, $\Delta$$\varepsilon$$^{ph}$$_{VBM}$ can be calculated using the finite displacement approach based on thermal lines \cite{S10,S11}. Unlike the conventional finite-displacement approach that evaluates each phonon separately and sum over all phonons, this stochastic approach considers all phonon modes at the same time, and the electron-phonon interaction can be calculated accurately and efficiently.

\section*{APPENDIX B: Derivation of $\Delta$\emph{E}$^{th}$}
Under the QHA, the thermal expansion induced energy correction \emph{E}$^{th}$, can be treated as arising from strain \cite{42,43}. Therefore, the corrections on the total energy differences between the system with and without defect $\alpha$ induced by the thermal expansion can be written as \cite{45}
\begin{equation}
\emph{E}^{\,th}(\alpha,\emph{q})[\emph{V}\,]-\emph{E}^{\,th}(host)[\emph{V}\,]=-2\gamma_0\,\Delta\emph{V}_{q}[\emph{V}-
\emph{V}_{0}(\emph{host})]+\Delta\gamma[\emph{V}-\emph{V}_{0}(\emph{host})]^2.   \tag{(B1)}
\end{equation}
\noindent $\gamma$$_0$ is the elastic constant of the host and $\Delta$$\gamma$ is the change of $\gamma$$_0$ induced by a defect $\alpha$ in charge state \emph{q}. $\Delta$\emph{V}$_{q}$=\emph{V}$_{0}$($\alpha$,\emph{q})-\emph{V}$_{0}$(host) is the volume change induced by the defect $\alpha$ at 0 K, in which \emph{V}$_{0}$(host) and \emph{V}$_{0}$($\alpha$,\emph{q}) are the equilibrium volume of the system without and with defect at 0K.

We assume that in a large system, a single defect cannot strongly influence the elastic constant, \emph{i.e.}, $\Delta$$\gamma$=0 as a first-order approximation, and $\Delta$\emph{E}$^{th}$ between the two charge states \emph{q} and \emph{q'} with the same defect $\alpha$ is
\begin{equation}
\Delta\emph{E}^{\,th}(\alpha,\emph{q}/\emph{q'})[\emph{V}\,]=\emph{E}^{\,th}(\alpha,\emph{q'})[\emph{V}\,]-
\emph{E}^{\,th}(\alpha,\emph{q})[\emph{V}\,]=-2\gamma_0\,[\emph{V}-\emph{V}_{0}(\emph{host})]\,\delta\emph{V}_{q\rightarrow q'},   \tag{(B2)}
\end{equation}
\noindent where $\delta$\emph{V}$_{q\rightarrow q'}$=\emph{V}$_{0}$($\alpha$,\emph{q'})-\emph{V}$_{0}$($\alpha$,\emph{q}) is the volume change induced by the defect $\alpha$ at 0 K when the charge-state changes from \emph{q} to \emph{q'}. Combining Eq.(A11), we have
\begin{equation}
\Delta\emph{E}^{\,th} =-2\gamma_0\,[\emph{V}-\emph{V}_{0}(\emph{host})]\,\delta\emph{V}_{q\rightarrow q'}
=-2\gamma_0\,\varphi_{V}\emph{T}\,\emph{V}_{0}(\emph{host})\,\delta\emph{V}_{q\rightarrow q'}.   \tag{(B3)}
\end{equation}

\section*{APPENDIX C: Derivation of $\Delta$\emph{F}$^{ph}$}
The phonon contribution to the free energy, \emph{F}$^{ph}$, can be described by Eq.(A9). Considering the first-order approximation of the Taylor expansion \emph{e$^x$=1+x+x$^2$/2!+x$^3$/3!+...}, we have
\begin{equation}
\emph{F}^{\,ph}[\emph{T}\,]= \sum_{i}[\frac{1}{2}\hbar\omega_{i}+\emph{k}_{B}\emph{T}\,\emph{ln}(\frac{\hbar \omega _i}{\emph{k}_B\emph{T}})].   \tag{(C1)}
\end{equation}

The $\Delta$\emph{F}$^{ph}$ between the two charge states \emph{q} and \emph{q'} with the same defect $\alpha$ is given by [two states with the same defect have the same number of phonon (\emph{i})]
\begin{align}
\Delta\emph{F}^{\,ph}(\alpha,\emph{q}/\emph{q'})[\emph{T}\,]=&\emph{F}^{\,ph}(\alpha,\emph{q'})[\emph{T}\,]-
\emph{F}^{\,ph}(\alpha,\emph{q})[\emph{T}\,]\nonumber \\
=&\sum_{i}[\frac{1}{2}\hbar\omega_{i}(\alpha,\emph{q'})+\emph{k}_{B}\emph{T}\,\emph{ln}(\frac{\hbar \omega _i (\alpha,\emph{q'})}{\emph{k}_B\emph{T}})]\nonumber \\
&-\sum_{i}[\frac{1}{2}\hbar\omega_{i}(\alpha,\emph{q})+
\emph{k}_{B}\emph{T}\,\emph{ln}(\frac{\hbar \omega _i (\alpha,\emph{q})}{\emph{k}_B\emph{T}})]\nonumber \\
=&\sum_{i}\lbrace\frac{1}{2}\hbar[\omega_{i}(\alpha,\emph{q'})-\omega_{i}(\alpha,\emph{q})]+
\emph{k}_{B}\emph{T}\,\emph{ln}[\frac{\omega_i(\alpha,\emph{q'})}{\omega_i(\alpha,\emph{q})}]\rbrace. \tag{(C2)}
\end{align}

Defining $\Delta$$\omega$$_{i}$= $\omega$$_{i}$($\alpha$,\emph{q'})-$\omega$$_{i}$($\alpha$,\emph{q}) and considering the first-order approximation of the Taylor expansion \emph{ln(1+x)=x-x$^2$/2+x$^3$/3-...}, we have
\begin{align}
\Delta\emph{F}^{\,ph}(\alpha,\emph{q}/\emph{q'})[\emph{T}\,]=&\sum_{i}\lbrace\frac{1}{2}\hbar\Delta\omega_{i}+
\emph{k}_{B}\emph{T}\,\emph{ln}[1+\frac{\Delta \omega_i}{\omega_i(\alpha,\emph{q})}]\rbrace\nonumber \\
=&\sum_{i}[\frac{1}{2}\hbar\Delta\omega_{i}+\emph{k}_{B}\emph{T}\,\frac{\Delta \omega _i}{\omega_i(\alpha,\emph{q})}].   \tag{(C3)}
\end{align}

\end{document}